# The Orientation Estimation of Elongated Underground Objects via Multi-Polarization Aggregation and Selection Neural Network

Hai-Han Sun, Yee Hui Lee, *Senior Member*, *IEEE*, Chongyi Li, Genevieve Ow, Mohamed Lokman Mohd Yusof, and Abdulkadir C. Yucel, *Senior Member*, *IEEE*

*Abstract*— **The horizontal orientation angle and vertical inclination angle of an elongated subsurface object are key parameters for object identification and imaging in ground penetrating radar (GPR) applications. Conventional methods can only extract the horizontal orientation angle or estimate both angles in narrow ranges due to limited polarimetric information and detection capability. To address these issues, this letter, for the first time, explores the possibility of leveraging neural networks with multi-polarimetric GPR data to estimate both angles of an elongated subsurface object in the entire spatial range. Based on the polarization-sensitive characteristic of an elongated object, we propose a multi-polarization aggregation and selection neural network (MASNet), which takes the multi-polarimetric radargrams as inputs, integrates their characteristics in the feature space, and selects discriminative features of reflected signal patterns for accurate orientation estimation. Numerical results show that our proposed MASNet achieves high estimation accuracy with an angle estimation error of less than 5°. The promising results obtained by the proposed method encourages one to think of new solutions for GPR-related tasks by integrating multi-polarization information with deep learning techniques. The data and code implemented in the paper can be found at https://haihan-sun.github.io/GPR.html.**

*Index Terms*— **Deep learning, ground-penetrating radar, multi-polarization, multi-task neural network, orientation estimation, elongated subsurface object**

## I. INTRODUCTION

GROUND penetrating radar (GPR) is a non-destructive method that uses electromagnetic waves to detect underground targets and map the subsurface environment [1]. When detecting elongated targets such as metal bars, unexploded ordnance, and tree roots, the information on the objects' horizontal and vertical orientations greatly improves the object classification accuracy [2] and facilitates the mapping of the underground environment [3].

Although elongated targets' orientation can be extracted from multiple parallel GPR B-scans [4], acquiring densely spaced scans is laborious and time-consuming. Besides, it may not be possible to conduct parallel scans over complex terrains. To estimate the objects' orientation, methods based on the polarization-sensitive characteristic [5]-[11] or the reflection pattern diversity [3] of an elongated object have been proposed.

Elongated targets depolarize the incident waves according to their orientation [5]. The received signal strength reflected by an elongated subsurface object is related to the object orientation and the polarization of the transmitting and receiving antennas (TX/RX) [6]. Therefore, multi-polarimetric antenna configurations have been utilized to improve GPR's ability to detect elongated objects and obtain information on their orientation [7]-[11]. The horizontal orientation of a buried elongated object can be extracted from the scattering matrix obtained by the multi-polarimetric antennas. In [9]-[11], the Alford rotation algorithm was applied to the multi-polarimetric scattering matrix to extract elongated objects' horizontal orientation by maximizing co-polarized or cross-polarized responses. In [6], the Frobenius norm of the matrix was used as a selection criterion of reflection amplitudes to ensure a stable angle estimation. In [2], the eigenvalues of the scattering matrix were calculated to find the horizontal orientation angle of the major axis of unexploded ordnance targets. To extract both the horizontal orientation angle and vertical inclination angle (in short, horizontal and vertical angles) of an underground elongated object, the study in [3] explored the relationship between the hyperbolic reflection shape and the parameters related to the object orientation angles by formulating a mathematical model.

The aforementioned conventional methods, however, fail to estimate both the horizontal and vertical angles in the entire spatial range. Their limitations are summarized as follows: 1) the Alford rotation algorithm [9]-[11] and the eigenvalue method [2] can only estimate the horizontal angle but not the vertical angle. This is because the polarization component perpendicular to the ground surface is difficult to obtain and not included in the GPR data. 2) The mathematical model proposed in [3] only estimates the horizontal and vertical angles in a constrained range of 0°-90° and 0°-45°, respectively. Besides, using only a single-polarized antenna for detection in [3] limits the GPR detection capability, thus resulting in indistinguishable reflection shapes in some cases and making it impossible to perform effective angle extraction. Therefore, it is still a challenge to effectively integrate multi-polarimetric GPR data to accurately and simultaneously estimate both horizontal and vertical angles of an elongated underground object.

Fortunately, powerful neural networks shed light on multi-modality data integration and parameter estimation through data-driven learning. Although neural networks have gradually been introduced into the electromagnetic inverse scattering problems in the GPR field, such as discriminating buried explosive targets [12], [13], estimating characteristics of objects [14], [15], and filtering antenna effects from GPR data [16], the combination of the multi-polarimetric GPR data and neural networks, to the best of our knowledge, has been never been reported before.

In this study, we explore the possible implementation of a neural network to integrate the information carried in multi-polarimetric scattering components for the accurate estimation of horizontal and vertical angles of an elongated subsurface object. Considering the complexity of the multi-polarimetric data, we specially design a novel Multi-Polarization Aggregation and Selection Network, called MASNet, for the angle estimation. The MASNet takes different combinations of multi-polarimetric data as input, integrates their information, extracts features using residual blocks, then adaptively selects distinguished features for angle estimation using convolutional layers and fully connected layers. The features are further analyzed by a multi-task branch structure to estimate both the horizontal and vertical angles simultaneously. The numerical experiment demonstrates that the trained MASNet can automatically and accurately estimate the orientation angles with a maximum error less than 5° in the entire spatial range, which outperforms the conventional methods. To the best of our knowledge, this work is the first that combines the multi-polarimetric radar domain knowledge with the learning capability of deep neural networks to solve a GPR-related task. The promising results obtained in this work make the proposed method and its variants highly suitable to solve other problems in polarimetric GPR applications.

## II. METHODOLOGY

In this section, we first discuss the response of different polarization components to elongated subsurface objects with different orientation angles, which motivates us to combine multi-polarization scattering components with the learning capability of neural networks to perform the orientation estimation task. Then, we introduce the MASNet, which is specially designed to simultaneously estimate both the horizontal and vertical angles. Finally, we present the loss function that is used during the optimization of the neural network.

### A. Response of Different Polarization Components to Subsurface Elongated Object with Different Orientation Angles

To illustrate different components' responses in detecting an underground elongated object with different orientation angles, we carried out a numerical study using gprMax [17]. The scenario used in this numerical study is shown in Fig. 1. A metal bar with a diameter of 0.6 cm and a length of 15 cm is buried in the sand. The distance from the metal bar's center to the soil surface is 28 mm. The object is oriented at a horizontal angle $\varphi$ and a vertical angle $\theta$, where $\varphi$ is the angle from the $x$-axis to the $y$-axis in the anti-clockwise direction on the $xy$-plane and $\theta$ is the angle from $x$-axis to the $z$-axis in the anti-clockwise direction. The sand has a relative permittivity of 4 and a conductivity of 0.001 S/m [18]. Orthogonally polarized ($x$- and $y$- polarized) source operating at 1.5 GHz and probe are used as transmitter and receiver to detect the object. The source and probe are separated 10 cm apart and placed above the sand surface at the height of 2 cm. The scanning is carried out by moving the source and probe along $x$-direction with a step size of 1.0 cm. 41 A-scans along the trace are collected. At each scanning position, the reflection from the object can be represented as a scattering matrix at a given time instant:

$$S = \begin{bmatrix} S_{xx} & S_{xy} \\ S_{yx} & S_{yy} \end{bmatrix}, \tag{1}$$

where the first and second subscripts in the matrix components represent the polarization of probe and source, respectively. Based on the reciprocity principle, $S_{xy}$ and $S_{yx}$ are equivalent in the homogenous sand environment. Therefore, three different polarimetric components $S_{xx}$, $S_{xy}$, and $S_{yy}$ are obtained. The GPR B-scans of these three components obtained after applying background removal as described in [19] are shown in Fig. 2.

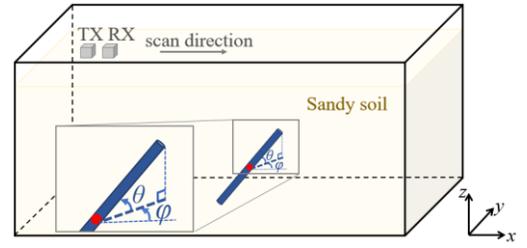

Fig. 1. Illustration of the simulation scenario in gprMax software.

Fig. 2(a) shows the B-scans of different scattering components when $\theta = 0°$ and $\varphi$ changes. When $\varphi = 0°$, $S_{xx}$ has the maximum value as the object direction is parallel to the $x$-direction. $S_{xy}$ and $S_{yy}$ have the minimum value. When $\varphi$ increases from 0° to 45°, $S_{xx}$ decreases, and $S_{xy}$ and $S_{yy}$ increase. At 45°, these three components have identical values as the reflected signal from the object has the same component in the $x$- and $y$-directions. When $\varphi$ further increases from 45° to 90°, $S_{yy}$ continues to increase, while $S_{xx}$ and $S_{xy}$ decrease. At 90°, $S_{yy}$ has the maximum value as the object direction is parallel to the $y$-direction, but $S_{xx}$ and $S_{xy}$ have the smallest values. When $\varphi$ increases from 90° to 135°, the variations of $S_{xx}$, $S_{xy}$, and $S_{yy}$ are opposite to the situation when $\varphi$ increases from 45° to 90°. At 135°, $S_{xx}$ and $S_{yy}$ are the same as the case at 45° as again the reflected signal has the same components in $x$- and $y$-directions. At this time, $S_{xy}$ can be used to distinguish two cases because $S_{xy}$ is out of phase in the two cases. In fact, the $S_{xx}$ and $S_{yy}$ are the same for the case ($\varphi_i, \theta_i$) and $(180°-\varphi_i, 180°-\theta_i)$, but $S_{xy}$ in these two cases are out of phase.

Besides the influence on the reflection intensity, the horizontal angle also affects the object's reflection shape. When $\varphi = 0°$ where the survey trace is parallel to the metal bar, $S_{xx}$ shows a flat reflection pattern as the distance between TX/RX and the bar is constant as GPR moves on the top of the bar.

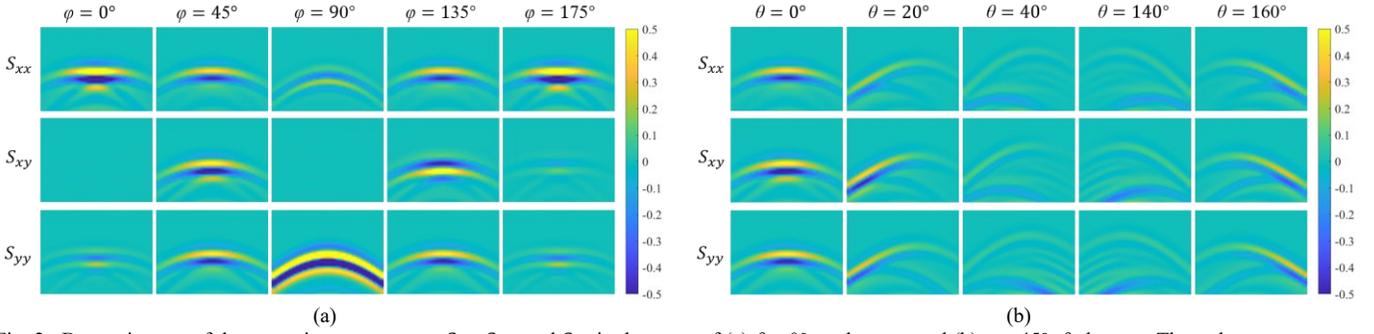

Fig. 2. B-scan images of the scattering components $S_{xx}$, $S_{xy}$, and $S_{yy}$ in the cases of (a) $\theta = 0°$, $\varphi$ changes, and (b) $\varphi = 45°$, $\theta$ changes. These three components enjoy different characteristics such as intensities and shapes, which provides the distinguished information for the accurate estimation of horizontal orientation angle and vertical inclination angle of an elongated subsurface object. B-scan images are represented by heatmaps, where different color indicates different values.

The edge reflections from the bar's two ends can also be observed. When $\varphi = 90°$, the survey trace is perpendicular to the direction of the bar, a well-formed hyperbolic curve for the $S_{yy}$ is observed. As $\varphi$ varies from 0° to 90°, the reflected signature changes from a flat pattern to a hyperbolic pattern, and as $\varphi$ varies from 90° to 180°, the trend is opposite.

Similarly, Fig. 2(b) shows the B-scans when $\varphi = 45°$, and $\theta$ changes. $\varphi = 45°$ is chosen because all the three components $S_{xx}$, $S_{xy}$, and $S_{yy}$ are distinguishable as indicated in Fig. 2(a). As shown in the figures, when $\theta$ increases from 0 to 90°, the hyperbolic shape representing the object's reflection gradually changes from symmetric to asymmetric, and the asymmetry becomes more obvious with a larger $\theta$. This is because for an elongated object with a vertical inclination angle $\theta$, one end of the object is closer to the soil surface while the other end is farther away from the surface, which results in different signal arrival times at two symmetrical positions above the object. When $\theta$ increases from 90° to 180°, the asymmetry appears in the other direction as the closest point is at the opposite end.

It can be concluded from the above analysis that (i) these three scattering components $S_{xx}$, $S_{xy}$, and $S_{yy}$ have different responses to object with different $\varphi$ and $\theta$, and (ii) the information they carry complement each other. Therefore, it is necessary to take this complementary information into account simultaneously to accurately estimate both the horizontal and vertical angles of an elongated object.

### B. Multi-Polarization Aggregation and Selection Network

To combine the domain knowledge of multi-polarization components with the learning capability of neural networks, we propose a novel multi-polarization aggregation and selection network, called MASNet, to estimate the horizontal and vertical angles of an elongated object. It is worth noting that there is no work that combines the multi-polarization GPR data with deep neural networks so far. The MASNet is specially designed to make full use of the complementarity and specificity of different polarization components for accurate estimation. The network structure of MASNet is shown in Fig. 3. To be specific, the MASNet includes three parts: 1) multi-polarimetric feature aggregation, 2) distinguished feature selection, and 3) multi-task angle estimation, which are detailed as follows.

**Multi-Polarimetric Feature Aggregation.** The inputs of the MASNet include the three individual scattering components

(i.e., $S_{xx}$, $S_{xy}$, and $S_{yy}$), and the concatenated groups along the channel dimension (i.e., $\{S_{xx}, S_{xy}\}, \{S_{xx}, S_{yy}\}, \{S_{xy}, S_{yy}\}$, and $\{S_{xx}, S_{xy}, S_{yy}\}$). Compared with only using the three individual components as inputs, the inclusion of the concatenated groups facilitates the network to learn complementary features from different polarization components and accelerates the network training convergence rate. Next, these groups of components are separately fed to the corresponding residual block [20] to extract independent features. The residual block contains three consecutive convolutional layers followed by the ReLU activation function. Each convolutional layer has 32 kernels of size 3×3 and stride 1. The residual block learns the residual function with reference to the layer input, which allows the network to learn identity-like mappings more easily. We found that the input images (i.e., three polarimetric radargrams) have small values. After passing through consecutive convolutional layers, these values may vanish, and thereby causing the collapse of the gradient backpropagation in the training phase. The skip-connection in the residual block can effectively remit this issue by learning the identity mapping.

**Distinguished Feature Selection.** From the multi-polarimetric features, we need to suppress the redundant information and select the more distinguished features for orientation estimation. To achieve that, we employ two convolutional layers followed by the ReLU activation function to shrink the number of feature maps from 32×7 to 32. Here, each convolutional layer has 32 kernels of size 3×3 and stride 1. It can be observed from the B-scans of scattering components for different object orientation angles in Fig. 2 that the difference of the reflection patterns appears in different spatial regions. Thus, to incorporate non-local information, we employ a fully connected layer with 1024 nodes to extract the distinguished features. We found that our simple structure already achieves good performance.

**Multi-Task Angle Estimation.** To estimate the horizontal and vertical angles simultaneously, we devise a multi-task branch structure. The selected features are forwarded to two parallel branches. Each branch contains two fully-connected layers for the angle estimation. Each fully connected layer consists of 1024 nodes.

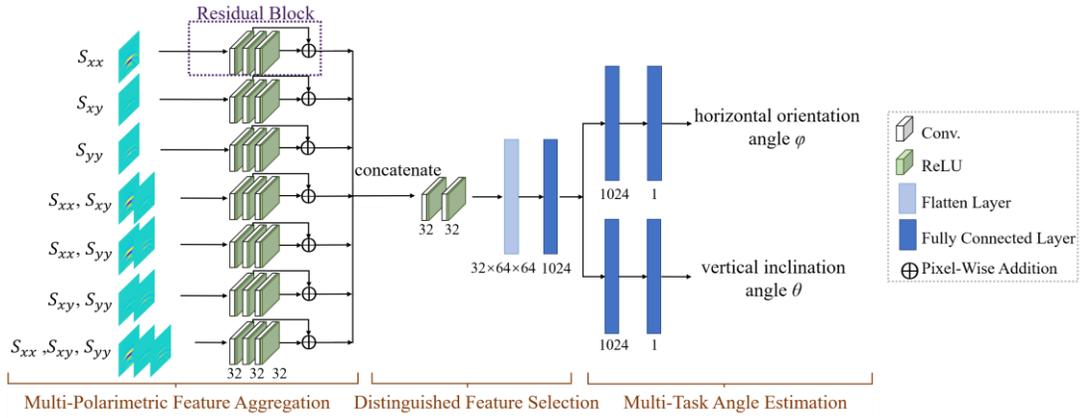

Fig. 3. The framework of the MASNet. The numbers shown in the figure indicate the numbers of output features (nodes).

## C. Loss Function

To drive the learning of our MASNet, we employ the mean squared error (MSE) as the loss function to optimize our network, which is expressed as:

$$MSE = \frac{1}{N}\sum_{i=1}^{N}((EstH_i - gtH_i)^2 + (EstV_i - gtV_i)^2), \quad (2)$$

where $N$ is the number of training data, $EstH$ and $EstV$ stand for the estimated horizontal and vertical angle, respectively. The corresponding ground truth is represented by $gtH$ and $gtV$.

## III. NUMERICAL RESULTS

### A. Implementation Details of the MASNet

To train the MASNet, we generated 1260 sets of B-scan images for a subsurface metal bar with different horizontal and vertical orientation angles $(\varphi, \theta)$ as described in Fig. 1. Each set contains the images of the scattering components $S_{xx}$, $S_{xy}$, and $S_{yy}$. Images are resized to 64×64 and normalized into [0, 1]. The angles $\varphi$ and $\theta$ both vary from 0 to 175° with a step size of 5°. We generated 60 sets of B-scan images for the metal bar with random orientation angles as the testing data. The testing images are also normalized and resized to 64×64.

We implement our framework with PyTorch on an NVIDIA 2080Ti GPU. The filter weights are initialized with the standard Gaussian function. We use the ADAM optimizer with default parameters and set the initial learning rate to 0.0001.

### B. Experimental results

The MASNet is applied to estimate orientation angles of the 60 sets of testing data. For illustration, 10 estimation results are listed in Table I. The estimation accuracy of the 60 estimation results is measured using mean absolute error (MAE) of the angles in degree. The MAE for $\varphi$ and $\theta$ are 1.2° and 0.9°, respectively. The maximum estimation error is less than 5°, proving the MASNet's capability in accurate 3-dimensional orientation estimation.

Additionally, we conduct ablation studies to investigate the effects of using different polarimetric components and their contribution to angle estimation. We retrain the ablated models while keeping the same settings as our final model (i.e., MASNet), except for the ablated parts. Table II lists the

### TABLE I
### THE ESTIMATED $(\varphi, \theta)$ FOR 10 EXAMPLES IN 60 TESTING SETS

| True angles | Estimated angles | True angles | Estimated angles |
|---|---|---|---|
| (16°, 24°) | (16.6°, 22.8°) | (24°, 118°) | (24.6°, 118.8°) |
| (30°, 34°) | (31.2°, 34.3°) | (50°, 20°) | (49.8°, 19.9°) |
| (63°, 175°) | (62.6°, 173.9°) | (85°, 133°) | (84.8°, 133.4°) |
| (96°, 131°) | (96.1°, 132.3°) | (110°, 44°) | (111.1°, 42.3°) |
| (135°, 74°) | (137.6°, 73.6°) | (171°, 45°) | (169.1°, 45.3°) |

### TABLE II
### COMPARISON OF ANGLE ESTIMATION ACCURACY IN THE ABLATION STUDY

| $S_{xx}$ | $S_{xy}$ | $S_{yy}$ | MAE ($\varphi$) | MAE ($\theta$) |
|---|---|---|---|---|
| √ | | | 37.2° | 52.3° |
| | √ | | 5.3° | 3.5° |
| | | √ | 37.6° | 52.4° |
| √ | √ | | 2.1° | 1.7° |
| √ | | √ | 37.3° | 52.3° |
| | √ | √ | 1.8° | 1.7° |
| √ | √ | √ | 1.2° | 0.9° |

### TABLE III
### COMPARISON OF ANGLE ESTIMATION ACCURACY ON TESTING SETS

| Methods | MAE ($\varphi$) | MAE ($\theta$) | Elapsed time |
|---|---|---|---|
| Alford rotation [6], [11] | 4.9° | - | 0.0063 s |
| Mathematical model [3] | 3.9° | 1.7° | non-automatic |
| The proposed MASNet | 1.2° | 0.9° | 0.0007 s |

quantitative comparisons of estimation accuracy on the testing sets. As presented, a network using only the $S_{xx}$ or $S_{yy}$ component or their combination yields high estimation error, as using two components cannot allow differentiating the combination for angle $(\varphi_i, \theta_i)$ from that of the angle $(180°-\varphi_i, 180° -\theta_i)$. A network using only the $S_{xy}$ component can improve the estimation accuracy as $S_{xy}$ contains information to differentiate the two intervals of $\varphi$. However, the improvement is constrained by the limited detection capability of $S_{xy}$. Combining $S_{xx}$ or $S_{yy}$ with $S_{xy}$ can significantly improve the estimation accuracy. This is because the neural network's estimation capability is enhanced by the complementary information of different polarimetric components. Our final model combining all three components achieves the lowest MAE, demonstrating the advantage of using multi-polarimetric data to improve the estimation accuracy.

To further demonstrate the advantages of the proposed method, we compare the angle estimation accuracy with the Alford rotation algorithm [6], [11] and the mathematical model [3]. We apply the Alford rotation algorithm with Frobenius norm $\geq 0.8$ as an amplitude filter in our testing data, and calculate the MAE as listed in Table III. The mathematical model presented in [3] is not suitable for the implementation in the entire angular range, thus we calculate the MAE using the reported simulation results with a 1600-MHz antenna in [3].

As presented in Table III, the proposed MASNet achieves the lowest MAE for both angles with the shortest elapsed time, which shows the effectiveness and efficiency of our proposed method in accurate angle estimation. In comparison, the Alford rotation algorithm can only estimate the horizontal angle $\varphi$ with limited accuracy, and it cannot extract the vertical angle. The mathematical model proposed in [3] can estimate both angles for infinite long objects within a limited angle range, but the estimation accuracy is slightly low for the horizontal angle, which may be caused by the use of single polarized data with limited detection capability. The comparison results clearly show that the proposed MASNet is capable of accurately estimating the orientation angles of an elongated object by effectively combining multi-polarimetric information and extracting the reflection pattern information.

A further numerical experiment was carried out to validate the proposed MASNet for angle estimation of elongated objects with different lengths and diameters. 9200 sets of multi-polarimetric B-scan data were randomly generated with the object's length and diameter ranging from 0.15 m to 0.9 m, and from 0.01 m to 0.04 m, respectively, while the horizontal and vertical angles vary from 0° to 179°. The survey trace is 0.4 m with its middle point located on the top of the object center. For an object with a length of more than 0.8 m that is much longer than the survey trace, it can be regarded as an infinite-length object. 9000 sets of data were used to train the MASNet, and 200 sets of data were used as the testing data. The mean absolute errors (MAE) of the horizontal and vertical angles estimated by the MASNet are 2.8° and 2.3°, respectively. The experiment demonstrates that the proposed MASNet maintains its estimation accuracy regardless of the length and diameter of the objects. Thus, in general, our method performs well for the angle estimation of various finite- and infinite-length elongated subsurface targets.

## IV. CONCLUSION

In this letter, we proposed a novel neural network, called MASNet, for the estimation of orientation angles of an elongated subsurface object. The network learns the relationship between the reflected signal patterns in the multi-polarimetric radargrams and the object orientation. Given multi-polarimetric radargrams, the network is capable of accurately estimating the orientation angles of the corresponding target. Compared with conventional methods, the proposed method covers the entire spatial angle ranges with high estimation accuracy.

The MASNet proposed in this letter is an attempt to combine the GPR domain knowledge with the learning capability of neural networks to estimate key characteristics of underground objects. The promising results obtained in the study can provide insights for the application of neural networks to the multi-polarimetric radargrams in the GPR field.